\documentclass[12pt]{article}
 \pdfoutput=1

\usepackage{amsmath,amssymb,amsfonts}
\usepackage{graphicx}
\usepackage{color}

\newcommand{\eref}[1]{Eq.~(\ref{e.#1})}

\newcommand{\cref}[1]{Chapter~\ref{c.#1}}

\newcommand{\barray}{\begin{eqnarray}}
\newcommand{\earray}{\end{eqnarray}}
\newcommand{\nn}{\nonumber \\}

\newcommand{\beq}{\begin{equation}}
\newcommand{\eeq}{\end{equation}}
\newcommand{\ba}{\begin{array}}
\newcommand{\ea}{\end{array}}
\newcommand{\bea}{\begin{eqnarray}}
\newcommand{\eea}{\end{eqnarray} }
\newcommand{\be}{\begin{eqnarray}}
\newcommand{\ee}{\end{eqnarray} }
\newcommand{\bal}{\begin{align}}
\newcommand{\eal}{\end{align}}
\newcommand{\bi}{\begin{itemize}}
\newcommand{\ei}{\end{itemize}}
\newcommand{\ben}{\begin{enumerate}}
\newcommand{\een}{\end{enumerate}}
\newcommand{\bc}{\begin{center}}
\newcommand{\ec}{\end{center}}
\newcommand{\bt}{\begin{table}}
\newcommand{\et}{\end{table}}
\newcommand{\btb}{\begin{tabular}}
\newcommand{\etb}{\end{tabular}}
\newcommand{\bvec}{\left ( \ba{c}}
\newcommand{\evec}{\ea \right )}

\def\cl{{\mathcal L}}

\def\co{{\mathcal O}}

\newcommand{\ti}{\tilde}
\def\hc{{\rm h.c.}}

\numberwithin{equation}{section}

\begin{document}
\begin{titlepage}
\vspace{-1cm}
\begin{flushright}
\small
LPT-ORSAY 12-14
\end{flushright}
\vspace{0.2cm}
\begin{center}
{\Large \bf Interpreting LHC Higgs Results \\ from Natural New Physics Perspective}
\vspace*{0.2cm}
\end{center}
\vskip0.2cm

\begin{center}
{\bf  Dean Carmi$^{a}$,  Adam Falkowski$^{b}$, Eric Kuflik$^{a}$, and Tomer Volansky$^{a}$}

\end{center}
\vskip 8pt

\begin{center}
{\it $^{a}$ Raymond and Beverly Sackler School of Physics and Astronomy, Tel-Aviv University, Tel-Aviv 69978, Israel}  \\
{\it $^{b}$ Laboratoire de Physique Th\'eorique d'Orsay, UMR8627--CNRS,\\ Universit\'e Paris--Sud, Orsay, France}
\end{center}

\vspace*{0.3cm}

\vglue 0.3truecm

\begin{abstract}
\vskip 3pt \noindent

We analyze the 2011 LHC and Tevatron Higgs data in the context of simplified new physics models addressing the naturalness problem.
These models are expected to contain new particles with sizable couplings to the Higgs boson, which can easily modify the Higgs production cross sections and branching fractions.
We  focus on searches in the $h\to ZZ^* \to 4l$, $h\to WW^* \to l\nu l \nu$, $h \to \gamma \gamma$, $h jj \to \gamma \gamma jj$ and $h V \to b \bar b V$ channels.
Combining the available ATLAS, CMS, and Tevatron data in these channels, we derive constraints on an effective low-energy theory of the Higgs boson. 
We then map several simplified scenarios to the effective theory, capturing numerous natural new physics models such as supersymmetry and Little Higgs,  and extract the constraints on the corresponding parameter space.
We show that simple models where one fermionic or one scalar partner is  responsible for stabilizing the Higgs potential are already constrained in a non-trivial way by  LHC and Tevatron Higgs data.

\end{abstract}

\end{titlepage}

\newpage

%%%%%%%%%%%%%%%
\section{Introduction}
%%%%%%%%%%%%%%

Discovering the Higgs boson and measuring its mass and branching ratios is one of the key objectives of the LHC.
Within the Standard Model (SM), the coupling to the Higgs boson is completely fixed by the mass of the particle.
This is no longer the case in many scenarios beyond the SM, where the Higgs couplings to the SM gauge bosons and fermions may display sizable departures from the SM predictions.
Indeed, precision studies of the Higgs couplings may be the shortest route to new physics.

Interestingly, from this point of view, a Higgs boson in the range $115-130$ GeV is particularly well suited as a new physics probe. 
One reason is that several different Higgs decay channels, in particular the $\gamma \gamma$, $ZZ^*$, $WW^*$, and $b\bar b$ channels, can be realistically accessed by experiment.  
The first of these arises in the SM at one loop and, consequently, physics beyond the SM may easily modify its rate.
This is especially true in models addressing the naturalness problem of electroweak symmetry breaking, which necessarily contain new charged particles with significant couplings to the Higgs boson. 
Well-known examples where this is the case include supersymmetric or composite Higgs models. 
Furthermore, the tree-level Higgs coupling to $WW$, $ZZ$, and $b\bar b$ is often modified as well, as is the case in composite or multi-Higgs models. 
Similar comments apply to the Higgs production rate: the dominant production mode via gluon fusion is a one loop process in the SM and is therefore particularly sensitive to new physics containing, as in typical natural models, light new colored states coupled to the Higgs.
Subleading production modes, such as vector boson fusion (VBF) and associate production, may also be affected.

While several Higgs production and decay modes may change in the presence of new particles, the correlated change in different channel may crucially depend on the new physics scenario. Consequently, a joint analysis of distinct independent channels may either allow to place interesting bounds on new physics scenarios or otherwise provide a way to discover new physics and pinpoint its identity.
The goal of this paper is to demonstrate the above understanding in light of the new Higgs measurements at the LHC and Tevatron, and place constraints on new physics models which solve the fine-tuning problem.

Recently, ATLAS, CMS, have reported the results of Higgs searches based on  5 fb$^{-1}$ of data  in several channels~\cite{Collaboration:2012si,Collaboration:2012sk,Collaboration:2012sm,Pieri:2012sb,Collaboration:2012tw, Chatrchyan:2012dg,Chatrchyan:2012ty,PAS_HIG,aad321}, and the Tevatron reported on the $b \bar b$ channel~\cite{TEVNPH:2012ab}.  The results, albeit inconclusive, suggest the existence of a Higgs boson with mass near 125 GeV.  It is therefore natural to try and answer the following question: {\it Assuming a Higgs boson with the mass $120 {\,\rm GeV} \leq m_h \leq 130 {\, \rm GeV}$,  what are the implications of these results for natural models beyond the SM?}  Below we pursue this question. 

We combine the latest ATLAS, CMS, and Tevatron Higgs results.
Our focus is to interpret the results in terms of simple (sometimes simplified) models that address the fine-tuning problem in the sense of providing a new contribution to the Higgs mass that cancels the quadratically divergent contribution of the SM top quark. 
To do so, we first consider the Higgs effective action at low energy and derive the constraints on its couplings.  We then map various theories onto the effective action to extract their bounds. 
A number of partly overlapping papers have recently investigated the 125 GeV Higgs-like excess in the context of composite Higgs~\cite{Espinosa:2012qj}, supersymmetric Higgs~\cite{Arbey:2011ab}, and multi-Higgs models \cite{Ferreira:2011aa}; see also \cite{all}.  For earlier related work, see \cite{Bonnet:2011yx}. 

Of course, at this stage the limited statistical power of the current Higgs data does not allow us to make a strong statement about any new physics scenario.  
Nevertheless, in several cases we are able to identify non-trivial regions of the parameter space that are disfavored at 95\% CL.
Repeating this analysis with future data may allow us, in the best case scenario, to pinpoint departures of the Higgs couplings from the SM predictions. 
That would not only provide evidence of new physics, but also some information about its scale, thereby supplying important hints about the nature of the fundamental theory at the electroweak scale.
 
The paper is organized as follows.  In the next section we define the effective action for a Higgs boson interacting with the SM fields and identify the relevant parameters that are being constrained by the present data. In Section~\ref{sec:constraints} we discuss the data and provide the combined best-fit of the ATLAS, CMS, and Tevatron Higgs results.  We then show the resulting constraints on the parameters of the Higgs effective action.  In Section~\ref{sec:scalar} we then study simplified models with scalar top partners, relevant for the MSSM as well.  Section~\ref{sec:fermion} focuses on fermionic top partners which show up in many composite Higgs  and Little Higgs models. Several representative examples are discussed.  Section~\ref{sec:multi-higgs} discusses some implications of theories with 2 Higgs particles.  We conclude in Section~\ref{sec:conclusions}.

%%%%%%%%%%%%%%%%%
\section{Formalism}
\label{sec:formalism}
%%%%%%%%%%%%%%%%%

We begin by defining a convenient framework to describe LHC and Tevatron Higgs phenomenology. 
We define an effective theory at the scale $\mu\sim m_h$, which describes the couplings of a single Higgs boson, $h$, to the SM gauge bosons and fermions.  Keeping dimension 5 operators and writing only couplings to the heaviest fermions we have,
\barray
\label{eq:1}
\cl_{eff} &= &
c_V {2 m_W^2 \over v} h \,  W_\mu^+ W_\mu^- + c_V  {m_Z^2 \over v} h \, Z_\mu Z_\mu - c_{b} {m_b \over v } h \,  \bar b b  - c_{\tau} {m_\tau \over v } h \, \bar \tau \tau
 \\ \nonumber &&
+ c_{g} {\alpha_s \over 12 \pi v} h \, G_{\mu \nu}^a G_{\mu \nu}^a + c_{\gamma } { \alpha \over \pi v} h \, A_{\mu \nu} A_{\mu \nu}\,.
\earray
Here $v = 246$ GeV, and $G_{\mu \nu}^a$ and $A_{\mu\nu}$ are the field strengths of the gluon and photon, respectively.
The fact that the same parameter $c_V$ controls the coupling to W and Z boson follows from the assumption that these couplings respect,  to a good approximation, custodial symmetry, as strongly suggested by electroweak precision observables.
We note that the Higgs could decay to particles from beyond the SM, e.g. to invisible collider-stable particles, but we will not discuss this possibility here. 
 We further note that while a single Higgs is kept at low energy, the above may describe multi-Higgs models, as long as there is a sizable splitting between the lightest and the remaining Higgs fields. We study such a possibility in more detail in Section~\ref{sec:multi-higgs}. 

In~\eqref{eq:1}, the top quark has been integrated out, contributing at 1-loop to $c_g$ and $c_\gamma$ as
\beq
\label{eq:cgtop}
c_g(\tau_t) = c_t A_f(\tau_t), \qquad c_\gamma(\tau_t) = {2 c_t \over 9} A_f(\tau_t), \qquad  A_f(\tau) = \frac{3}{2\tau^2} \left [ (\tau-1)f(\tau) + \tau \right ],
\eeq
where $\tau_t = m_h^2/4m_t^2$, $c_t$ is the ratio of the top-Higgs Yukawa coupling to the SM one,  and
\begin{eqnarray}
\label{eq:3}
f(\tau)
= \left\{ \begin{array}{lll}
{\rm arcsin}^2\sqrt{\tau} && \tau \le 1 \\ -\frac{1}{4}\left[\log\frac{1+\sqrt{1-\tau^{-1}}}{1-\sqrt{1-\tau^{-1}}}-i\pi\right]^2 && \tau > 1 \end{array}\right.\,.
\end{eqnarray}
For $m_h^2 \ll 4 m_t^2$ one finds,  $f(\tau) \simeq \tau(1+ \tau/3)$, which is a very good approximation for $m_h \lesssim 130$ GeV.
Consequently, for the SM with a light Higgs boson matched to our effective theory at 1-loop we have 
\begin{eqnarray}
\label{eq:2}
c_{V,\rm SM} = c_{b,\rm SM} =1\,,\qquad\qquad c_{g,\rm SM} \simeq 1\,, \qquad\qquad c_{\gamma,\rm SM} \simeq 2/9\,.
\end{eqnarray}

The decay widths of the Higgs relative to the SM predictions are modified approximately as,
\bea
{\Gamma(h \to b \bar b) \over \Gamma_{SM}(h \to b \bar b) } = |c_b|^2 \,,
&\qquad &
{\Gamma(h \to WW^*) \over \Gamma_{SM}(h \to W W^*) } = {\Gamma(h \to ZZ^*) \over \Gamma_{SM}(h \to ZZ^*) } = |c_V|^2 \,,
 \nn
{\Gamma(h \to gg) \over \Gamma_{SM}(h \to gg) } \simeq |c_g|^2 \,,
&\qquad &
{\Gamma(h \to \gamma \gamma) \over \Gamma_{SM}(h \to \gamma \gamma) } = \left |\hat c_{\gamma} \over \hat c_{\gamma,SM} \right|^2 \,,
\earray
where $\hat c_\gamma$ includes also the one-loop contribution due to the triangle diagram with the W boson\footnote{There are additional one-loop contributions to $c_
\gamma$ and $c_g$ from light quarks that are left out in this discussion, but are included in the analyses below.},
\beq
\hat c_\gamma(\tau_t, \tau_W) = c_\gamma(\tau_t) - \frac{c_V}{8\tau_W^2}\left[3(2\tau_W-1)f(\tau_W)+3\tau_W+2\tau_W^2\right]\,,
 \eeq
where $\tau_W = m_h^2/4m_W^2$. For $m_h = 125$ GeV one finds $\hat c_{\gamma} \simeq c_\gamma - 1.04 c_V$, and thus $ \hat c_{\gamma,SM} \simeq - 0.81$.

More generally, the 1-loop contribution to $c_g$ from an additional fermion in the fundamental representation of $SU(3)_C$ and coupled to the Higgs via the Yukawa coupling $y_f h \bar f /\sqrt{2}$ is simply given by Eq.~\eqref{eq:cgtop}  with 
$c_t \to (v y_f/\sqrt 2 m_f)$ and $\tau_t \to \tau_f$, while for an $SU(3)_C$ fundamental scalar,
\beq 
\label{eq:cg}
\delta c_g(\tau_s)= \frac{1}{4} \sum_s g_{hss} A_{s}(\tau_s) \,, \qquad g_{hs s} = \frac{1}{2}\frac{v}{m_s^2}\frac{\partial m_s^2}{\partial v}\,, 
\qquad A_{s}(\tau) = \frac{3}{\tau^2} \left [ f(\tau) - \tau \right ]\,. 
\eeq 
For the photon coupling we have 
\begin{eqnarray}
\label{eq:cgamma}
\delta c_\gamma(\tau_{f,s}) =  {Q_{f,s}^2 \over 2} \delta c_g (\tau_{f,s}), 
\end{eqnarray}
where $Q_{f,s}$ is the electric charge of the scalar or fermion. 
More general expressions can be found e.g in \cite{Djouadi:2005gi}. 
Note that in the limit $\tau \rightarrow 0$, $A_{f,s}(\tau)\rightarrow 1$, and the scalar contribution becomes $1/4$ that of the fermion. 
In fact, the $\tau \rightarrow 0$ limit is equivalent to approximating $c_g$ and $c_\gamma$ using the 1-loop beta function \cite{Ellis:1975ap}, which explains the relative factor $1/4$.

As discussed in the introduction, the most significant constraints on the effective theory are obtained by studying several independent Higgs decay channels. The five most constraining channels  to date are $h\to ZZ^*$, $h\to WW^*$, $h\to \gamma\gamma$, $pp \to hV\to b \bar b V$ and $pp \to hjj\to \gamma\gamma jj$.  The Higgs production mechanism in the first three channels is dominated by the gluon fusion process which scales as $c_g^2$. The $b \bar b$ channel is dominated by associate production which scales as $c_V^2$. Thus, the relevant Higgs event rates scale as,
\begin{eqnarray}
\label{eq:4}
R_{VV} &\equiv& {\sigma(pp \to h) {\rm Br}(h \to VV^*) \over \sigma_{SM}(pp \to h) {\rm Br}_{SM}(h \to VV^*) } \simeq  \left|{c_g c_V \over C_{tot}}\right|^2  \, , 
\\
\label{eq:5}
R_{\gamma \gamma} &\equiv& {\sigma(pp \to h) {\rm Br}(h \to \gamma \gamma ) \over \sigma_{SM}(pp \to h) {\rm Br}_{SM}(h \to \gamma \gamma) } \simeq  \left| {c_g \hat c_{\gamma} \over \hat c_{\gamma,SM} C_{tot}}\right|^2\,,
\\
R_{b\bar{b}V} &\equiv& {\sigma(p\bar p \to hV) {\rm Br}(h \to b \bar b  ) \over \sigma_{SM}(p \bar p \to hV) {\rm Br}_{SM}(h \to b \bar b) } \simeq \left|{c_V c_b  \over   C_{tot}}\right|^2\,.
\end{eqnarray}
where $ |C_{tot}|^2 = \Gamma_{tot} / \Gamma_{tot}^{SM}$. The approximation holds assuming the Higgs production remains dominated by the gluon fusion subprocess. More precise relations are used in our fits. 

The $\gamma \gamma jj$ channel is slightly more complex, since it receives comparable contributions from the gluon fusion and VBF production channels:
\begin{eqnarray}
\label{eq:6}
R_{\gamma \gamma jj} &\equiv& {\sigma(pp \to hjj) {\rm Br}(h \to \gamma \gamma ) \over \sigma_{SM}(pp \to hjj) {\rm Br}_{SM}(h \to \gamma \gamma) }=  \frac{\sigma^{SM}_{ggf}\cdot \epsilon_{ggf}\cdot |c_g|^2+\sigma^{SM}_{VBF}\cdot \epsilon_{VBF}\cdot |c_V|^2}{\sigma^{SM}_{ggf} \cdot \epsilon_{ggf}+\sigma^{SM}_{VBF}\cdot \epsilon_{VBF}} \cdot \left|{\hat c_{\gamma} \over \hat c_{\gamma,SM} C_{tot}}\right|^2 
\nonumber\\
\end{eqnarray}
where $\epsilon_{ggf}$ and $\epsilon_{VBF}$ are the efficiencies to pass the selection cuts in the gluon fusion and VBF production modes, respectively.  In the SM, the gluon fusion mode contributes about $1/3$ that of the VBF production mode to the $2 \gamma 2j$ final state studied by CMS  \cite{Collaboration:2012tw}, but it may become more important in models where the gluon fusion cross section is enhanced relative to the VBF one.\footnote{We thank Yevgeny Kats for pointing this out to us. }

%Plugging the SM cross sections in the last equation gives:
%\begin{eqnarray}
%\label{eq:666}
%R_{\gamma \gamma jj} \sim \frac{15.3\epsilon_{ggf}\cdot c_g^2+1.2\epsilon_{VBF}\cdot c_V^2}{15.3\epsilon_{ggf}+1.2\epsilon_{VBF}} \cdot \left|{\hat c_{\gamma} \over \hat c_{\gamma,SM} C_{tot}}\right|^2 
%\end{eqnarray}

%In the above equations we used:
%\bea
%|C_{tot}|^2 \equiv \frac{\Gamma_{tot}}{\Gamma^{SM}_{tot}}= 0.577 |c_b|^2+0.241 |c_V|^2 + \ldots
%\eea
%The Higgs  rate is the same in the $WW$ and $ZZ$ channels as a consequence of  custodial symmetry. 
%In the next sections, we construct the bounds on these rates.

%%%%%%%%%%%%%%%%%
\section{Constraints from the LHC and Tevatron}
\label{sec:constraints} 
%%%%%%%%%%%%%%%%%

\begin{figure}[!]
\vspace{1cm}
\bc
\includegraphics[width=\textwidth]{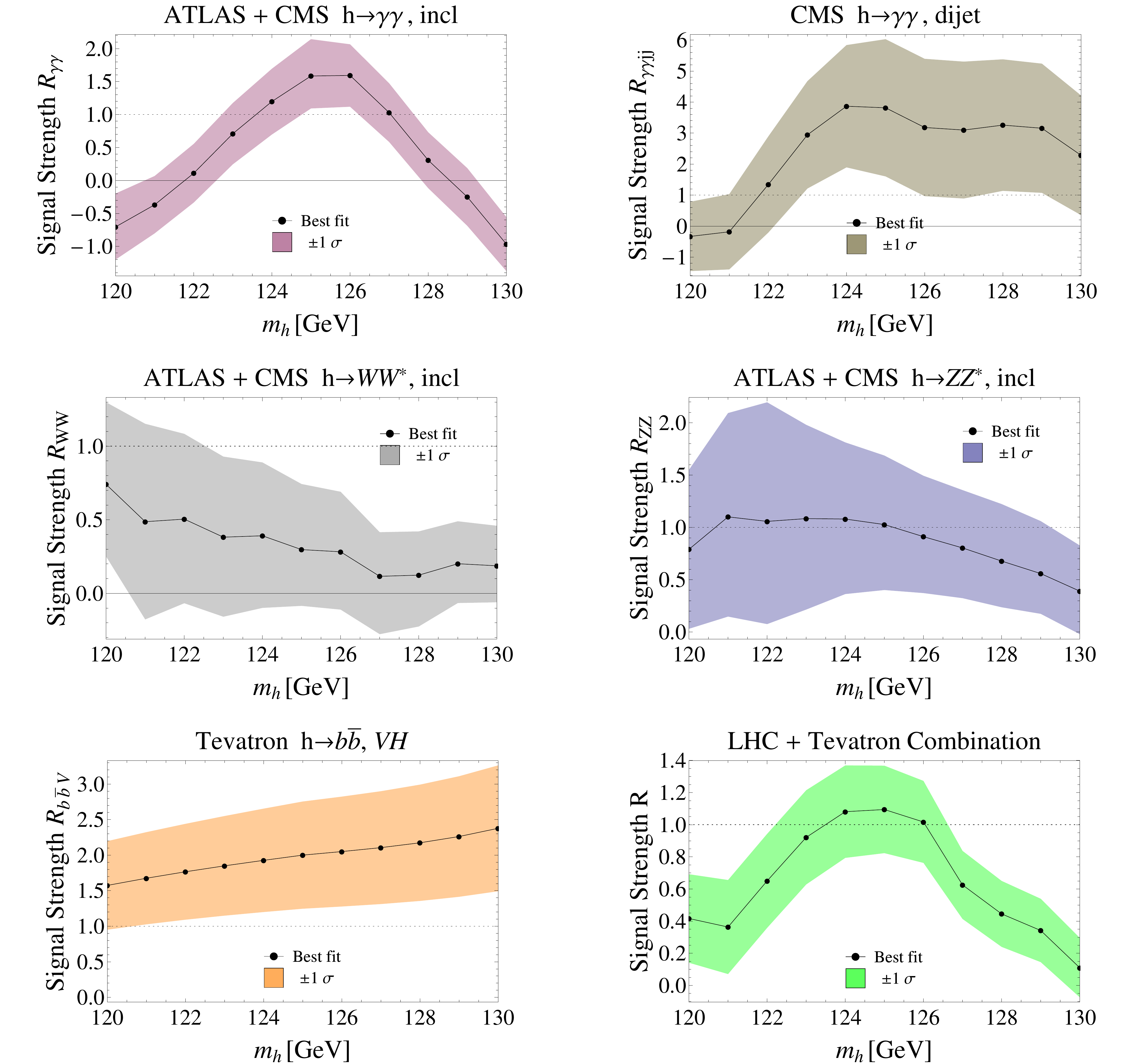}
\ec
\vspace{0mm}
\caption{\small Best-fit values and $1\sigma$ bands for the rates $R_{\gamma \gamma}$, $R_{\gamma \gamma jj}$,  $R_{WW}$, $R_{ZZ}$, and $R_{b \bar b}$ for a Higgs mass between 120 GeV and 130 GeV. We also show the combination of all the channels ({\bf bottom-right}).   In all but the CMS dijet measurement, the results are computed using the reported results and assuming gaussian statistics.   For the CMS dijet, the best fits are derived by repeating the analysis reported in~\cite{Collaboration:2012tw}, not taking  systematic uncertainties into account.   The results in the dijet mode are found to be conservative.
 \label{fig:muhat}}
\end{figure}

Recently, LHC and Tevatron have reported the results of Higgs searches in several channels. 
Here we focus on the following channels: $h\to \gamma \gamma$ and $pp\to hjj \to \gamma \gamma jj$\  \cite{Collaboration:2012sk,Collaboration:2012tw}, $h \to ZZ^* \to 4l$ \cite{Collaboration:2012sm,Chatrchyan:2012dg},  $h \to WW^* \to 4l$ \cite{aad321},  $Vh \to V b \bar  b$  \cite{TEVNPH:2012ab}  channels, which are currently the most sensitive ones for $115 < m_h < 130$ GeV. 

Both LHC and Tevatron observe an excess of events that (inconclusively) indicate the existence of a Higgs boson with mass in the $124-126$ GeV range.
The largest excess comes from CMS in the rate $R_{\gamma \gamma jj}$, and hints towards a cross-section which is larger then in the SM. In the diphoton channel, both CMS and ATLAS observe a rate $R_{\gamma \gamma}$ which is consistent with the SM. The Tevatron sees a small excess in $R_{b \bar b}$, as compared to the SM higgs. Consequently, as will be shown below, combining the results together points to a production cross-section and branching fractions consistent  with that predicted by the SM.  It remains to be seen whether with better statistics and improved understanding of the systematics, the results will remain consistent with the SM prediction, or otherwise converge on a rate deviating from that predicted by the SM.

\begin{figure}[!]
\vspace{0cm}
\bc
\includegraphics[width=0.9\textwidth]{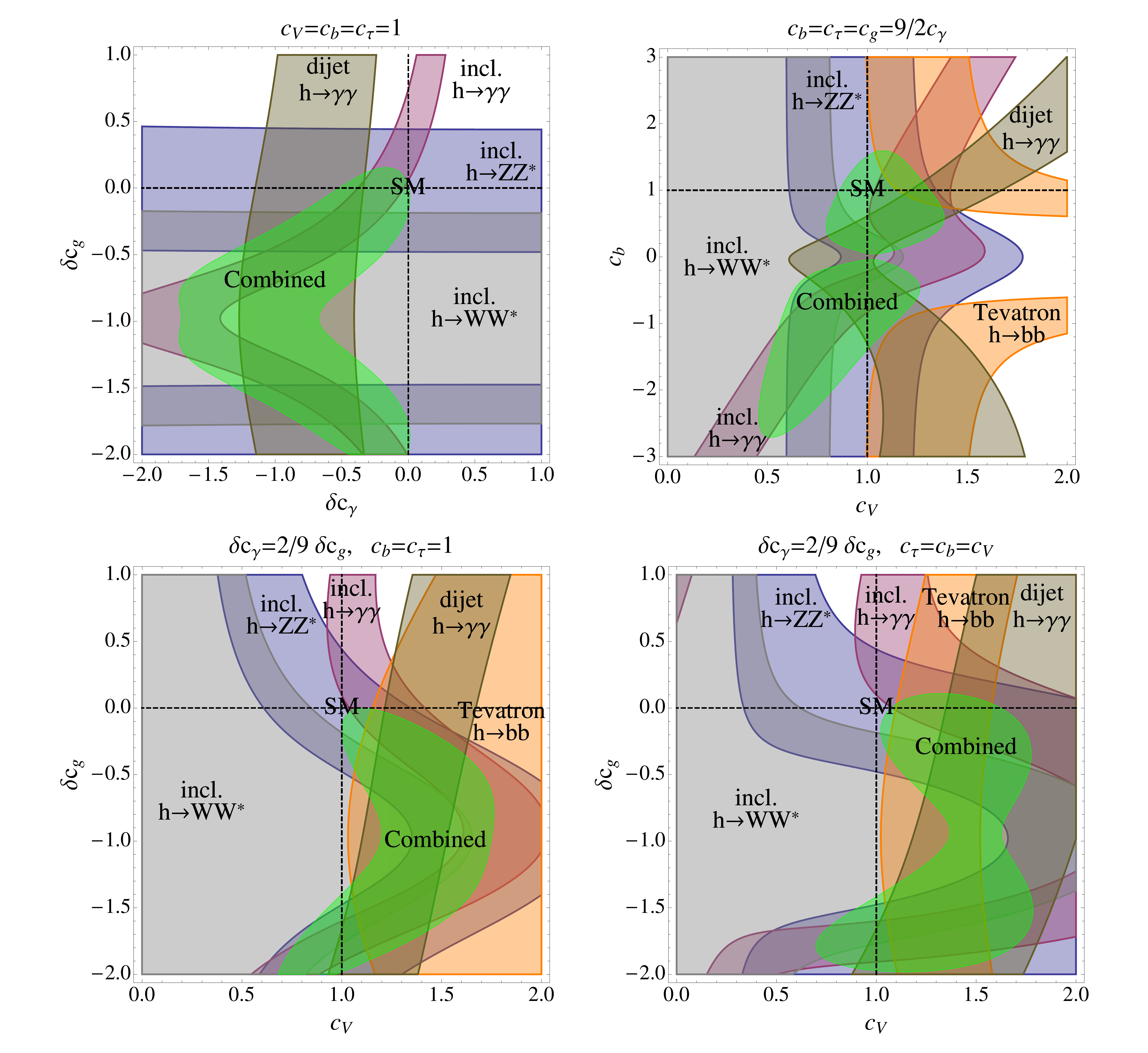}
\ec
\vspace{-1cm}
\caption{\small 
The allowed parameter space of the effective theory given in Eq.~\eqref{eq:1}, derived from the LHC and Tevatron constraints for $m_h = 125$ GeV. 
We display the $1\sigma$ allowed regions for the rates in  Eqs.~\eqref{eq:4}-\eqref{eq:6}: $R_{\gamma \gamma}$ (purple), $R_{ZZ}$ (blue), $R_{WW}$ (light grey), $R_{\gamma \gamma jj}$ (beige), and $R_{b \bar b}$ (orange). The ``Combined'' region (green) shows the 95\% CL allowed region arising from all channels. The crossing of the dashed lines is the SM point. The top left plot characterizes models in which loops containing beyond the SM fields contribute to the effective 5-dimensional $h \, G_{\mu \nu}^a G_{\mu \nu}^a$ and $h \, A_{\mu \nu} A_{\mu \nu}$ operators, while leaving the lower-dimension Higgs couplings in Eq.~\eqref{eq:1} unchanged relative to the SM prediction. The top right plot characterizes composite Higgs models and can be compared to \cite{Azatov:2012bz} and \cite{Espinosa:2012ir}. The lower  plots characterize {\emph{top partner}} models where  only scalars and fermions with the same charge and color as the top quark contribute to the effective 5-dimensional operators, which implies the relation $\delta c_\gamma = (2/9) \delta c_g$.  
The results are shown for 2 different sets of assumptions about the lower-dimension Higgs couplings that can be realized in concrete models addressing the Higgs naturalness problem.  
%The top right plot was added in v2 to allow a direct comparison with the results of  Refs.  
 \label{fig:generalFit}}
\end{figure}

In order to constrain the couplings of the effective theory -- $c_g, c_\gamma, c_V, c_b$ and $c_\tau$ in Eq.~\eqref{eq:1} -- it is crucial to analyze several Higgs production and decay modes. Following the discussion above, we focus on the channels which are the most constraining, i.e Eqs.~\eqref{eq:4}-\eqref{eq:6}. Since one of the production modes (gluon fusion) and one of the decay modes ($\gamma\gamma$) are loop-induced, these constraints are very sensitive to heavy particles beyond the SM that may play a role in solving the fine tuning problem, leading to interesting conclusions on new physics and naturalness.

 In Fig.~\ref{fig:muhat} we show the results of the combined best fit value of $\hat R  \equiv \sigma/\sigma_{\rm SM}$, 
 assuming gaussian statistics,
  for each of the analyzed channels separately and for the combination of all channels, for Higgs mass between 120 GeV and 130 GeV.  The bands indicate the $1\sigma$ uncertainty. Since the CMS experiment  does not provide the values of $R_{\gamma \gamma j j}$ for $m_H=120-130$ GeV, we calculated the best fit for the rates in the channel, which we show in Fig.~\ref{fig:muhat}.
More specifically, we repeat the analyses, computing the likelihood functions.  We use background and signal modeling given by the experiments, normalizing the signal to the reported values. %In the remaining cases we simulate the signal using Madgraph~\cite{Alwall:2011uj} and Delphes~\cite{Ovyn:2009tx}, and implement the reported cuts. 
For the results shown here, we do not take into account the systematic effects which are expected to be significant in the dijet channel of the diphoton analysis. 

  %For $m_h \approx 125$ GeV, which we find best explains the combined measurements, the best fit cross-section is consistent with the SM cross-section.

 %In all of our results, our estimation of the 1-sigma band for $\hat\mu$ is larger than that reported by the individual experiments. From this point of view, our results are expected to be conservative and the constraints based on the prospective \textcolor{red}{official ATLAS/CMS combination} may be stronger than those reported here. Our procedure does not (and cannot) take into account sone important information and should therefore be taken with a grain of salt.

In Fig.~\ref{fig:generalFit}  we use our results of Fig.~\ref{fig:muhat} to place constraints on the effective theory assuming $m_h=125$ GeV.  We show two dimensional constraints on $\delta c_g = c_g - c_{g,\rm SM}$, $\delta c_\gamma = c_\gamma - c_{\gamma,\rm SM}$ and $c_V$ for various model assumptions.  Shown are the $1\sigma$ allowed regions for $R_{\gamma \gamma}$ (purple), $R_{ZZ}$ (blue), $R_{WW}$ (light grey), $R_{\gamma \gamma jj}$ (beige), $R_{b \bar b}$ (orange). The green region gives the allowed region at $95\%$ CL for the combination of all channels.  In Fig~\ref{fig:generalFit}a we allow only the Higgs couplings to gluons and photons to change while keeping the other couplings at the SM values. In the remaining plots of Fig.~\ref{fig:generalFit} we keep $\delta c_\gamma/\delta c_g = 2/9$ fixed, while varying the other couplings. That  ratio is  conserved when top partners with  the same  charge and color as the top are introduced.

In the next three sections, we study various models that allow for an improvement in the fine-tuning of the Higgs mass.  Our goal is to keep the discussion quite general, and we therefore consider simplified models that capture different paradigms showing up in many models that solve the fine-tuning problem. Each of the models is then mapped on to the effective theory, Eq.~\eqref{eq:1}, and the constraints on the Higgs rate in various channels derived above are used to place bounds on specific scenarios. Throughout the paper we use the constraints  depicted in Fig.~\ref{fig:muhat}, except for the plots assuming $m_h =125$ GeV for which the bounds on $R_{\gamma \gamma jj}$ are taken from  the CMS MVA analysis in~\cite{Collaboration:2012tw} .

%%%%%%%%%%%%%%%%%%%%%%%%%%%%%%
\section{Models with Scalar Top Partners}
\label{sec:scalar}
%%%%%%%%%%%%%%%%%%%%%%%%%%%%%%

%----------------------------------
\subsection{One Scalar}
\label{sec:onscalar}

\begin{figure}[tb]
\vspace{0cm}
\bc
\includegraphics[width=1\textwidth]{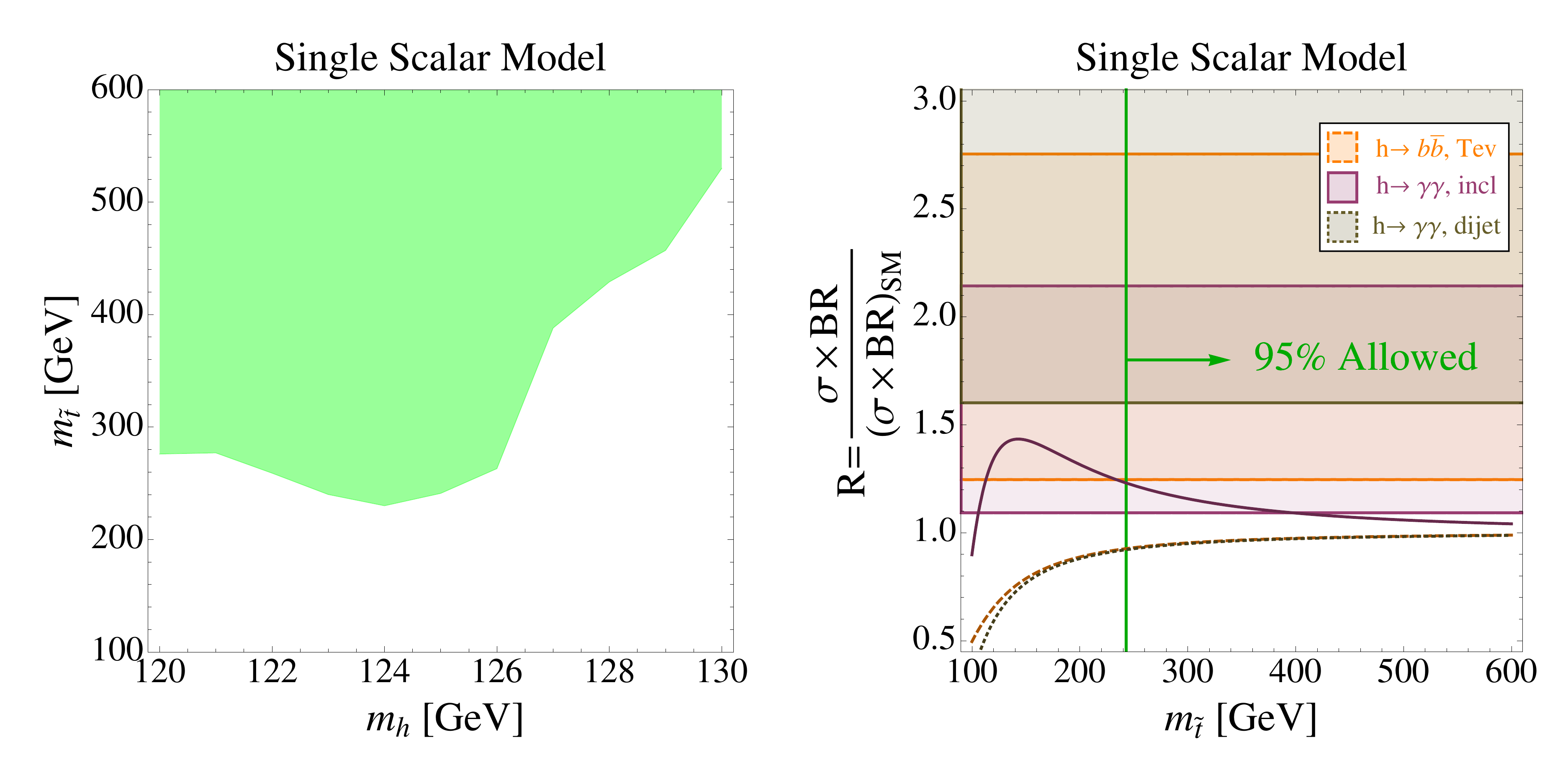}
\ec
\vspace{-1cm}
\caption{\small{\textbf{Left:} Favored region, 95\% CL, in the $m_{\tilde{t}}-m_h$ plane, derived from the combination of all search channels, for the one-scalar model described in Sec.~\ref{sec:onscalar}. 
\textbf{Right:} Constraints for $m_h=125$ GeV. The three bands show the $1\sigma$ allowed regions: $R_{\gamma \gamma}$ (purple), $R_{b \bar b}$ (orange), $R_{\gamma \gamma jj}$ (beige). The three curves show the theoretical predictions as a function of  $m_{\tilde{t}}$:  $R_{\gamma \gamma}$  (solid-purple),  $R_{b \bar b}$ (dashed-orange), and  $R_{\gamma \gamma jj}$ (dotted-beige). Only 3 channels are shown, but all channels are included. The region to the right of the green line at $m_{\tilde{t}}= 240$ GeV shows the 95\% CL experimental allowed region. }
\label{fig:SingleScalarR}
}
\end{figure}

We start our exploration  with the simple toy model of a single scalar top partner.
Consider  a scalar $\tilde t$  with electric charge 2/3 and transforming in the fundamental representation under the $SU(3)$ color. 
At the renormalizable level, the top sector mass and interaction terms can be parametrized as
\beq
\label{e.1s}
\cl_{stop} =   - \left ( y H Q t^c  + \hc   \right ) -   |\ti t|^2 \left ( M^2  + \lambda  |H|^2 \right ). 
\eeq
Here $Q = (t,b)$ is the 3rd generation quark doublet, $t^c$ is the  $SU(2)_W$  singlet  top and $H$ is the Higgs doublet.  In the unitary gauge $H = (0, (v + h)/\sqrt{2})$ and $|H| =  (v+ h)/\sqrt{2}$, where  $v = 246$ GeV and  $h$ is the canonically normalized Higgs boson field.
It follows that $m_{\tilde t}^2 = M^2 + \lambda v^2/2$.
The quadratic divergent top contribution to  the Higgs mass is canceled  by the scalar partner when the  coupling $\lambda$ is related to the top Yukawa coupling by
\beq
\label{e.DivergenceConstraintScalar}
\lambda  = 2 y^2 .
\eeq
Note this is different than in minimal supersymmetry where 2 scalar partners with $\lambda \simeq y^2$ play a role in canceling the top quadratic divergence. 

For $m_{\tilde t} \gg m_h/2$,  using Eqs.~\eqref{eq:cg},\eqref{eq:cgamma} one finds the scalar partner contribution to the effective dimension 5 operator,
\beq
{c_g \over c_{g,\rm SM}}  =  {c_\gamma \over c_{\gamma,\rm SM}}    \simeq  1 + \lambda {v^2 \over 8 m_{\tilde t }^2} = 1 + {m_t^2 \over 2 m_{\tilde t}^2}\,, \qquad\qquad {c_V=c_b=1}\,.
\eeq
The last equality holds when   \eref{DivergenceConstraintScalar} is satisfied.
Thus, if the scalar top partner is soley responsible for the cancellation of the top quadratic divergence, then the gluon fusion rate is always {\em enhanced}, while the diphoton rate  is slightly suppressed for realistic $m_{\tilde t}$ (due to interference with the negative W loop contribution). This is unlike the MSSM where  both enhancement and suppression of the gluon fusion can be realized within the realistic  parameter space (see below).

In Fig. \ref{fig:SingleScalarR} we show the 95\% CL allowed region for $m_{\tilde t}$ as a function of the Higgs mass (left), along with the $1\sigma$ bounds for $m_h = 125$ GeV (right).    We see that model independently, a single scalar top partner lighter than  $240$ GeV is excluded, if indeed the LHC and Tevatron signals correspond to a 125 GeV Higgs boson, as hinted by the data.

%----------------------------------
\subsection{Two Scalars (MSSM)}

\begin{figure}[tb]
\vspace{0cm}
\bc
\includegraphics[width=1\textwidth]{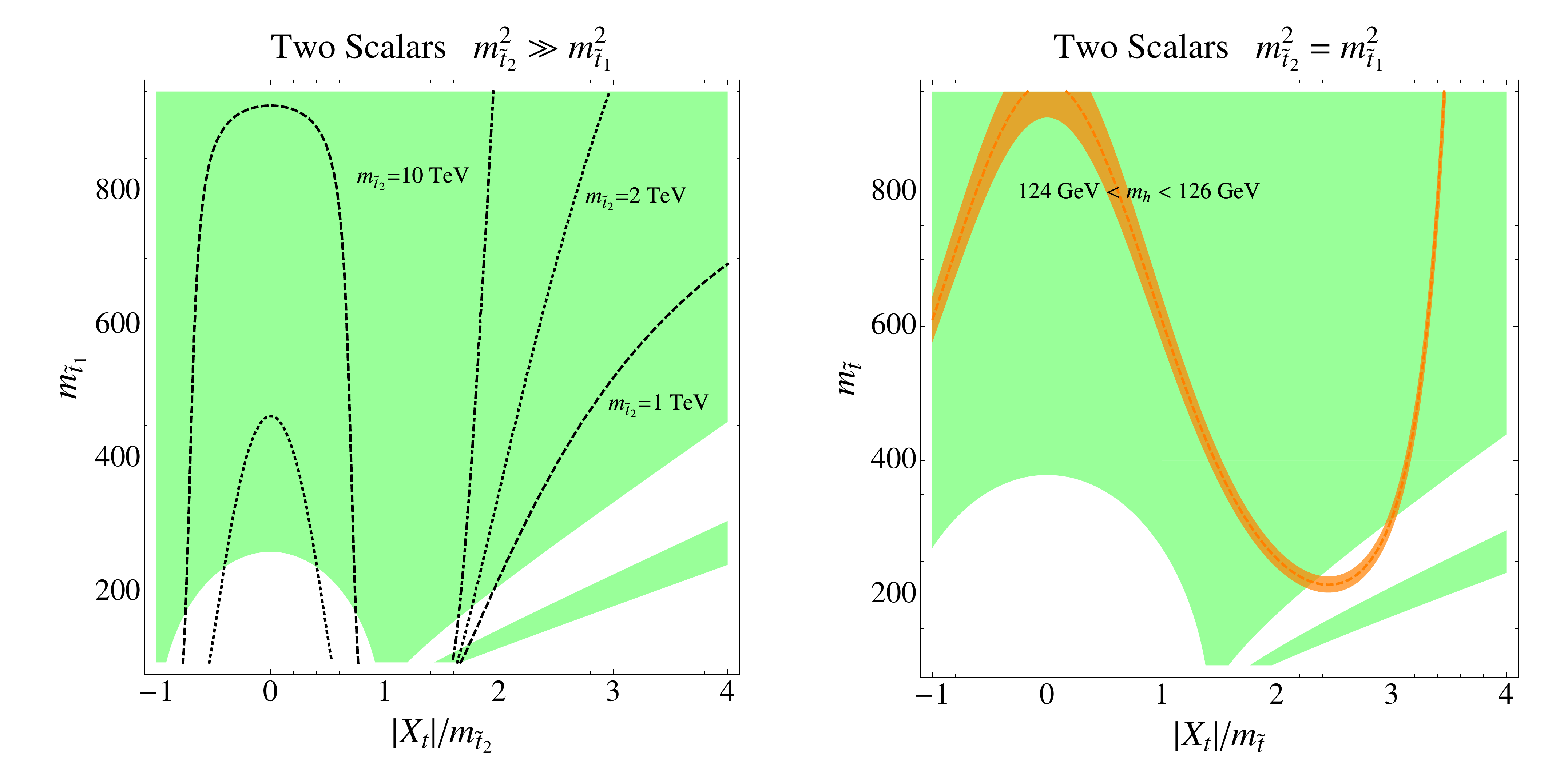}
\ec
\vspace{-.5cm}
\caption{\small\textbf{Left:} The favored region at 95\% CL for $m_h = 125$ GeV, derived from the combination of all search channels, in the two scalar model with $m_{{\tilde{t}_2}} \gg m_{{\tilde{t}_1}}$. Also shown are contours of constant $m_h = 125$ GeV assuming the 1-loop MSSM relation between  Higgs and stop masses, for  $m_{\tilde{t}_2}=1,2,$ and 10  TeV.
\textbf{Right:} Same for  $m_{{\tilde{t}_2}} = m_{{\tilde{t}_1}}$. Also shown is a band corresponding to 124 GeV $<m_h<$ 16 GeV  assuming the 1-loop MSSM relation between  Higgs and stop masses.  Additional, model-dependent, bounds on stops from direct searches are not shown.   
 \label{fig:SUSY}
}
\end{figure}

Consider the system of 2 scalar top partners $\ti t,\ti t^c$, one for the left-handed top and one for the right-handed top, with the mass terms of the form
\beq
-\cl_{stop} =   |\ti t|^2 \left ( \ti m^2  + y^2 |H|^2 \right )  +   |\ti t^c|^2 \left ( \ti m_c^2  + y^2 |H|^2 \right )
+  y |H| X_t \left ( \ti t \ti t^c + \hc \right),
\eeq
where $y$ is the top Yukawa coupling, as in \eref{1s}.
This is equivalent to the stop sector of the MSSM in  the decoupling limit ($M_A \gg M_Z$) and neglecting the  (sub-leading)  D-terms contribution to the stop masses.
Here the contributions of both scalars {sum} to cancel the quadratic divergence from the top quark. 
The left-handed and right-handed stops mix in the presence of $X_t$, which in the MSSM  is given by  $X_t = |A_t - \mu \cot\beta|$. See e.g \cite{Dermisek:2007fi}.

Denoting the two mass eigenvalues by $m_{\tilde{t}_{i}}$, and the left-right mixing angle by $\theta_t$, one has
\beq
m_t X_t = \frac{1}{2}\left( m_{\tilde{t}_2}^2 - m_{\tilde{t}_1}^2 \right) \sin 2 \theta_t
\eeq
where, by convention,  $m_{\tilde{t}_1} \le m_{\tilde{t}_2}$.
For $m_{\tilde{t}_{i}} \gg m_h/2$, integrating out the stops shifts the effective dimension-5 operators as
\beq
\label{e.SUSYc}
{c_g \over c_{g,\rm SM}}  =  {c_\gamma \over c_{\gamma,\rm SM}}    =
1 + {1 \over 4} \left({m_t^2 \over  m_{\tilde{t}_1}^2}   + {m_t^2
\over m_{\tilde{t}_2}^2} - {m_t^2 X_t^2\over  m_{\tilde{t}_1}^2 m_{\tilde{t}_2}^2} \right)
\eeq
For zero mixing, the stops always  interfere constructively with the top contribution (destructively with the $W$-contribution to $\hat{c}_\gamma$), but once $X_t$ becomes comparable to stop masses an enhancement of $c_g$ becomes possible. A significant shift of the gluon fusion  and diphoton widths is possible if at least one of the stop mass eigenvalues is close to the top mass, or if the mixing is very large.

In Fig.~\ref{fig:SUSY}  we illustrate the impact of the LHC and Tevatron Higgs data on the parameter space of the 2-scalar-partner model.
The left  plot shows the allowed region in the $m_{\tilde{t}_1} $--$X_t / m_{\tilde{t}_2} $ plane, assuming that $m_{\tilde{t}_2}^2$
is large enough so that the heavier stop eigenstate does not contribute to the effective operators (that is, dropping the second term in the bracket in \eref{SUSYc}).
For no mixing,  $X_t^2 / m_{\tilde{t}_2}^2 =0$, the lower bound on the lightest stop is $\sim 250$ GeV.  %\textcolor{red}{This bound is weaker than the} one in the single partner model because the coupling of each stop is a factor of 2 smaller.   On the right plot we consider the case of degenerate stops. Here, for no mixing, we get the same bound on the stop mass as in  the single partner model.     
 
In both scenarios, for just right amount of mixing, that is $X_t/m_{\tilde{t}_2} \simeq 1$ for $m_{\tilde{t}_2}^2 \gg
m_{\tilde{t}_1}^2$ and $|X_t |/ m_{\tilde{t}_2} \simeq \sqrt{2}$ for $m_{\tilde{t}_2}^2 = m_{\tilde{t}_1}^2$, the scalar partners  contribution to
$c_g$ and $c_\gamma$ can vanish, even for very light stops. 
This may be relevant for models that require a light stop, such as electroweak baryogenesis \cite{Carena:2008vj}. 
% Interestingly, in the first case, this can also maximize the 1-loop contribution to the Higgs mass in the MSSM.
For illustration, on the left plot of Fig.~\ref{fig:SUSY} we show contours of constant $m_h = 125$ GeV, for $m_{\tilde{t}_2}=1,2,$ and 10  TeV, while on
the right plot we show the region where 124 GeV $< m_h <$ 126 GeV.
We note that that  used the one-loop formula for the Higgs mass in the MSSM, therefore these contours should be considered illustrative only.
As a final remark, we comment that additional bounds on stops exist from direct searches. 
These bounds are however model dependent, in particular strongly  depending on the stop decay branching fractions,  and therefore we do not display them.

%%%%%%%%%%%%%%%%%%%%%%%%%%%%%%%%%
\section{Models with a Fermionic Top Partner}
\label{sec:fermion}
%%%%%%%%%%%%%%%%%%%%%%%%%%%%%%%%%

We move to the case of one {fermionic} top partner.
Consider the SM model extended by a vector-like quark pair $(T,T^c)$ in the ${\bf 1}_{2/3}$ representation under $SU(2)_W \times U(1)_Y$.
Fermionic  partners cannot cancel the top quadratic divergence if the effective Lagrangian describing    their interactions with the Higgs is renormalizable.
Therefore in this case we need to consider a more general effective Lagrangian for the top sector that includes non-renormalizable interactions,
\bea
\label{e.TopParametrization}
- \cl_{top} &=&  y_1(|H|^2) H Q t^c +  y_2 (|H|^2) H Q T^c + M_1(|H|^2)  T t^c +  M_2(|H|^2)  T T^c    + \hc \,.
\eea
We allow the vacuum expectation value  of the Higgs doublet,  $\hat v$, to be different from the electroweak scale $v = 246$ GeV, which may happen if the Higgs effective interactions with W/Z bosons are also non-renormalizable and corresponds to $c_V \neq 1$ (this is in fact the case in Little Higgs and composite Higgs models).
We assume that all mass and Yukawa couplings are  functions of $|H|^2$ and can be expanded in powers of $|H|^2/M^2$ where $M$ is the mass scale of the heavy top quark.
Up to order $|H|^2/M^2$ they can be parametrized as
\bea
%\begin{array}{lclcclcl}
y_1(|H|^2) &=&  y_1 \left (1  -  d_1 {|H|^2 \over M^2} \right)  + \co \Big(\frac{|H|^4}{M^4}\Big)\,,
\\
y_2(|H|^2) &=& y_2 \left (1  -  d_2{|H|^2 \over  M^2} \right)  + \co \Big(\frac{|H|^4}{M^4}\Big)\,,
\\
M_1 (|H|^2) &=&  c_1 M  {|H|^2 \over M^2} + \co \Big(\frac{|H|^4}{M^4}\Big)\,,
\\
M_2 (|H|^2) &=&  M \left (1 -  c_2  {|H|^2 \over M^2}  \right)+ \co \Big(\frac{|H|^4}{M^4}\Big)\,.
%\end{array}
\eea
Above, we used the freedom to rotate  $t^c$ and  $T^c$ such that $M_1$ starts at  $\co(|H|^2)$.
In terms of these parameters  $m_{top} \simeq y_1 \hat v /\sqrt{2}$ while $m_T \simeq M$.
For the cancellation of the quadratic divergences in the Higgs mass term, one straightforwardly finds,
\beq
\label{e.DivergenceConstraintFermionic}
c_2 = {y_1^2 + y_2^2  \over 2}\,.
\eeq
This relation may arise naturally in models where the Higgs is realized as a pseudo-Goldstone boson of a spontaneously broken approximate global symmetry.

Following the discussion above Eq.~\eqref{eq:cg} and integrating out the top sector, one finds for the effective Higgs coupling to gluons and photons shifts as
 \beq
\label{e.Rtops}
  {c_g \over c_{g,\rm SM}}  =  {c_\gamma \over c_{\gamma,\rm SM}}
 \simeq  {v \over \hat v} \left [ 1  -   {\hat v^2   \over M^2} \left ( d_1 +   c_2  +  {c_1 y_2 \over y_1} \right)  \right ]\,.
\eeq
We see that several parameters of the effective Lagrangian enter the modification of effective Higgs coupling to gluons and photons.
Above, $y_1$ can be eliminated in favor of the top mass, and $c_2$ can be eliminated using the condition \eref{DivergenceConstraintFermionic}.
This still leaves 4 free parameters: $d_1$, $y_2$, $\hat v/v$ and $M$.
Thus, in full generality, we cannot predict the magnitude, or even the sign of the correction to the Higgs rate merely by demanding cancellation of quadratic divergences.\footnote{In composite Higgs models under certain conditions one can argue that the gluon fusion and diphoton decay rate cannot be enhanced \cite{Low:2009di}.}
However, concrete realizations of Little Higgs and composite Higgs models often  imply additional relations between the effective theory parameters, in which case the set-up becomes more predictive.
Below we study several  predictive patterns of effective theory parameters that arise in popular Little Higgs and composite Higgs models.

%-----------------------------------------------------
\subsection{No mixing}
\label{sec:nomixing}

 \begin{figure}[!]
\bc
\includegraphics[width=1\textwidth]{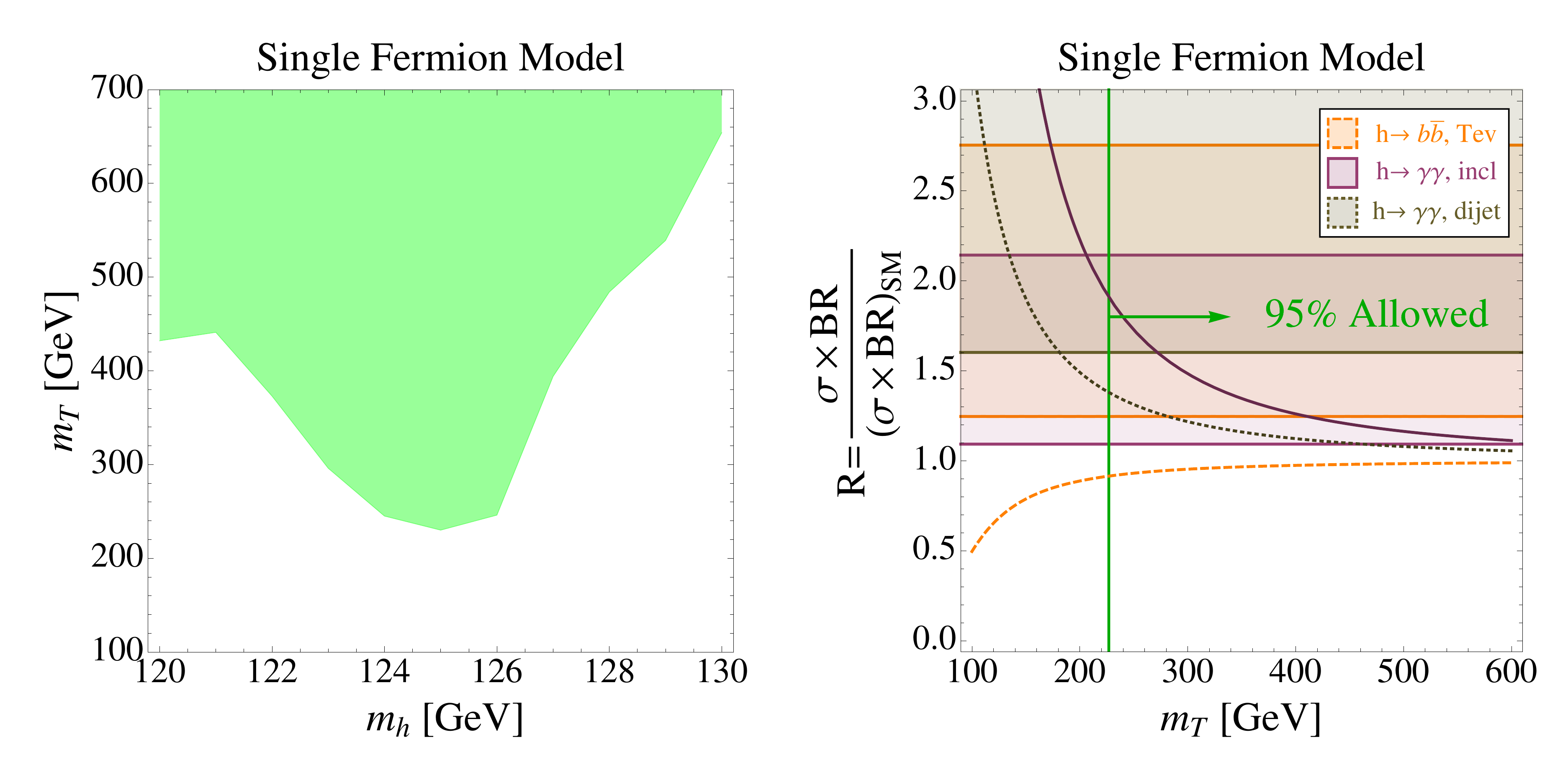}
\ec
\vspace{-1cm}
\caption{\small{\textbf{Left:} Favored region, 95\% CL, in the $m_{T}-m_h$ plane, derived from the combination of all search channels, for the single-fermion, no-mixing model described in Sec.~\ref{sec:nomixing}. 
\textbf{Right:} Constraints assuming $m_h=125$ GeV. The three bands show the $1\sigma$ allowed regions: $R_{\gamma \gamma}$ (purple), $R_{b \bar b}$ (orange), $R_{\gamma \gamma jj}$ (beige). The three curves show the theoretical predictions as a function of  $m_T$: $R_{\gamma \gamma}$ (solid-purple), $R_{b \bar b}$ (dashed-orange) and $R_{\gamma \gamma jj}$ (dotted-beige). Only 3 channels are shown, but all channels are included. The region to the right of the green line at $m_T= 220$ GeV shows the 95\% CL experimental (combined) allowed region. }
 \label{fig:SingleFermionR}}
\end{figure}

First, we will restrict the parameter space by demanding that the SM top does not mix with its partners, $c_1 = y_2 = 0$, and $c_2 = y_1^2/2 \simeq
m_t^2/v^2$.
This situation occurs in Little Higgs with T-parity \cite{Cheng:2003ju}.
Furthermore, we assume that the Higgs coupling to the SM fields is not modified at $\co(v^3)$, thus $\hat v = v$ and $d_1 = 0$.
Under these assumptions one finds
\beq
{c_g \over c_{g,\rm SM}}  =  {c_\gamma \over c_{\gamma,\rm SM}}
 \simeq  1  -  { m_t^2 \over m_T^2}\,,
 \qquad c_V = c_b = 1\,.
\eeq
Hence in this scenario, much as in the case of one scalar partner,  the departure of the Higgs rates from the SM can be described by one parameter:  the ratio of the top mass to its partner mass.
The gluon width, and in consequence the dominant Higgs production mode, is reduced.
On the other hand, the Higgs partial width into $WW$ and $ZZ$ are unchanged, while the partial width in the  $\gamma \gamma$ channel is significantly changed only when $m_T \sim m_t$.
In Fig.~\ref{fig:SingleFermionR} we present the constraints on $m_T$ from the LHC and Tevatron Higgs data.  In the left plot we show the 95\% CL allowed region, in the $m_h$-$m_T$ plane.   
The right plot shows the constraints assuming a 125 GeV Higgs.  As can be seen, $m_T \lesssim 220$ GeV is disfavored in this case.

%-----------------------------------------------------
\subsection{Universal suppression}
\label{s.slh}

Consider now a frequently occurring situation when all the Higgs rates are suppressed by a universal factor depending on  the compositeness scale $f$.
To be specific, consider the top sector interacting with a pseudo-Goldstone Higgs as
\beq
\label{e.topSLH}
-\cl_{top} = y f \sin(|H|/f) t  t^c  +  y f \cos(|H|/f) T  t^c  + M' T T^c +\hc  .
\eeq
The top partner  mass is of order $m_T \simeq \sqrt{y^2 f^2 + M'^2}$.
Integrating out the top sector we find,
\beq
{c_g \over c_{g,\rm SM}}  =  {c_\gamma \over c_{\gamma,\rm SM}}   
= \cos (\hat v/\sqrt 2 f) =  \sqrt{1 - {v^2 \over 2 f^2} }\,.
\eeq
Thus, the top sector contribution to the Higgs dimension-5 interactions is reduced by a factor that is independent of the details of the top sector, such as the masses and the coupling of the top eigenstates.
The interaction terms in \eref{topSLH} arise e.g. in  the Simplest Little Higgs model  with an $[SU(3)/SU(2)]^2$ coset structure \cite{Schmaltz:2004de} when taking the limit $f_2 \gg f_1$.
In that case one also finds $c_V = c_b = \sqrt{1 - {v^2 \over 2 f^2} }$.
Therefore, in the Simplest Little Higgs model, the rates in all Higgs channels  are universally {suppressed} by a factor depending only on the compositeness scale: $\sigma/\sigma_{SM} = 1 - v^2/2f^2$. The same holds  for the $SO(5)/SO(4)$ minimal composite Higgs with fermions embedded in the spinorial representation of $SO(5)$ \cite{Agashe:2004rs}.
Note that the independence of the Higgs widths of the fine details of the top sector  persists in numerous Little Higgs and composite Higgs models  \cite{Falkowski:2007hz},   although it may not hold in more complicated models where the top couples to more than one composite operator \cite{Azatov:2011qy}.

\begin{figure}[!]
\bc
\includegraphics[width=1\textwidth]{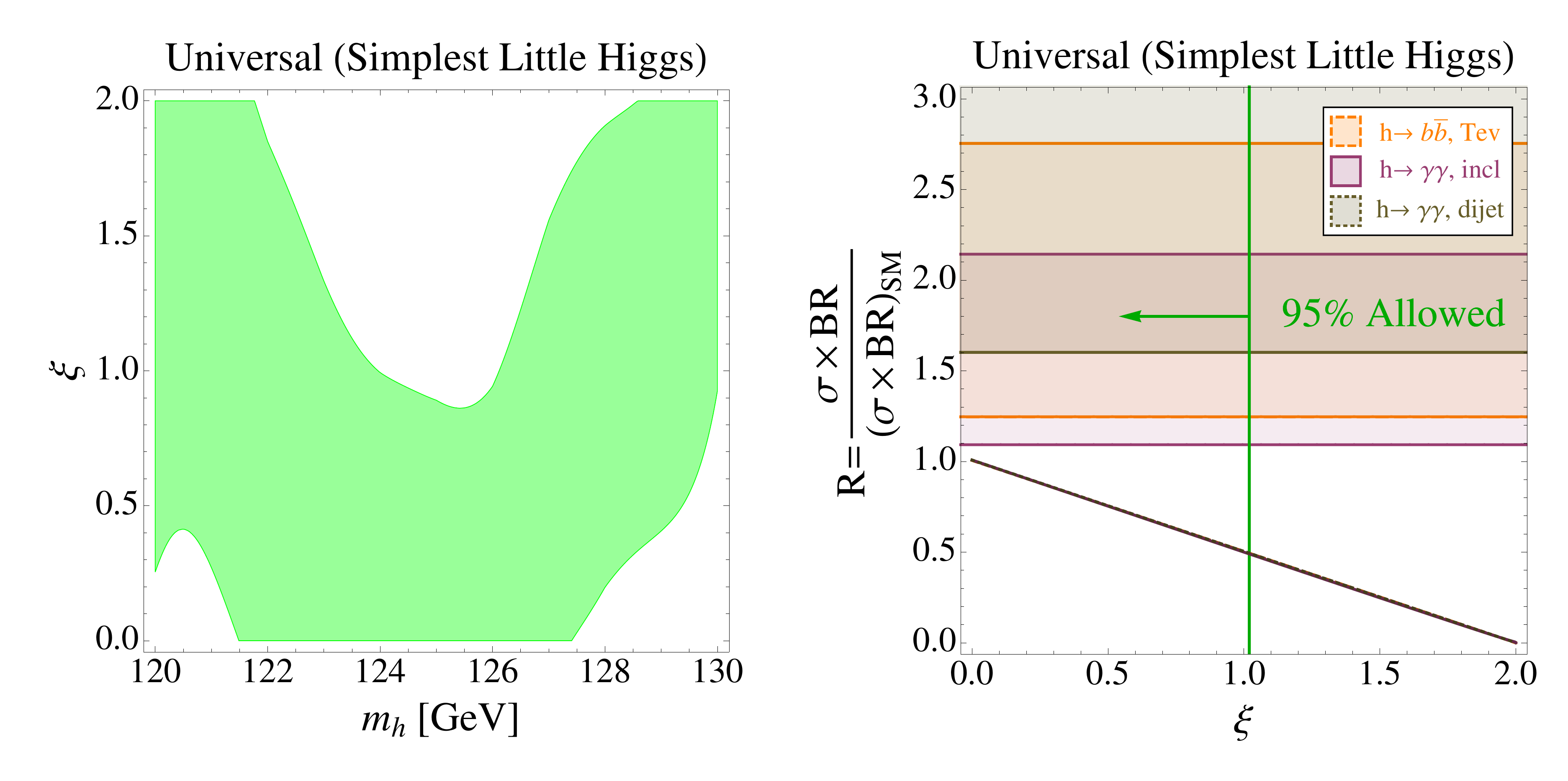}
\ec
\vspace{-1cm}
\caption{\small{ \textbf{Left:} Favored region, 95\% CL, in the $\xi -m_h$ plane where $\xi \equiv \frac{v^2}{f^2}$, derived from the combination of all search channels, for models with universal suppression such as the Simplest Little Higgs model described in Sec.~\ref{s.slh}. 
\textbf{Right:} Constraints for $m_h=125$ GeV.  The three bands show the $1\sigma$ allowed regions:  $R_{\gamma \gamma}$ (purple), $R_{b \bar b}$ (orange), $R_{\gamma \gamma jj}$ (beige). The three curves show the theoretical predictions as a function of  $\xi$: $R_{\gamma \gamma}$ (solid-purple), $R_{b \bar b}$  (dashed-orange) and $R_{\gamma \gamma jj}$ (dotted-beige). Only 3 channels are shown, but all channels are included. Due to the universal suppression all three curves share the same dependence on $\xi$ and are therefore on top of one another.  The region to the left of the green line at $\xi \simeq 1$ shows the 95\% CL experimental (combined) allowed region.  }  
 \label{fig:SimplestHiggsR}}
\end{figure}

Repeating the analysis done in previous sections, in Fig.~\ref{fig:SimplestHiggsR} we present the constraints on $\xi \equiv v^2/f^2$ from the current LHC and Tevatron Higgs measurements.   We find that, assuming a 125 GeV Higgs boson, $\xi>1$ is excluded at the 95\% CL.    Note that as discussed above, all relative rates have similar dependence on $\xi$ and are therefore  drawn on top of one another.

%-----------------------------------------------------
\subsection{Non-universal suppression}
%-------------------------------------------------------
\label{sec:twinhiggs}

\begin{figure}[tb]
\bc
\includegraphics[width=1\textwidth]{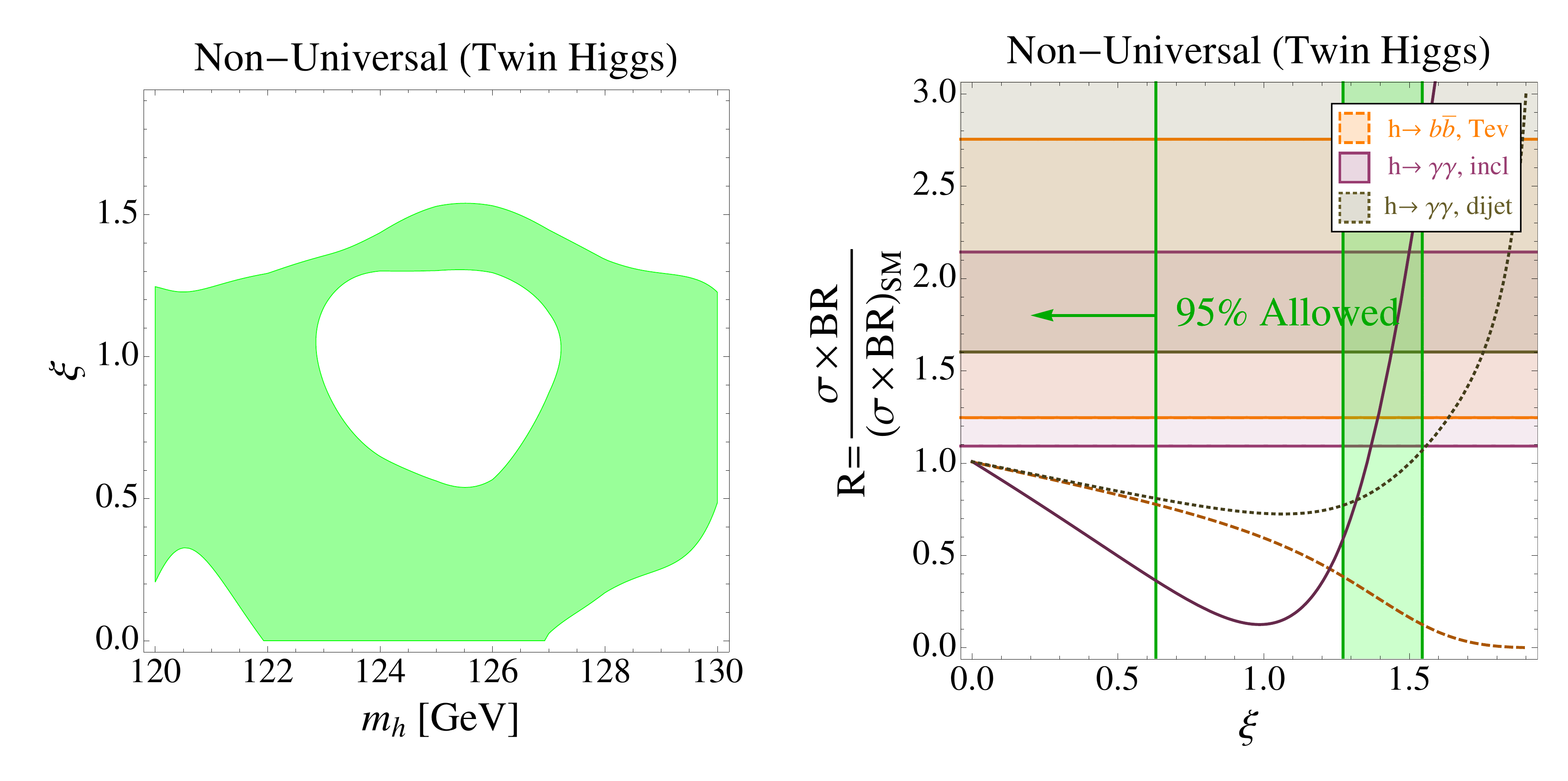} 
\ec
\vspace{-1cm}
\caption{\small{\textbf{Left:} Favored region, 95\% CL, in the $\xi \equiv \frac{v^2}{f^2} -m_h$ plane, derived from the combination of all search channels, for the Twin Higgs model described in Sec.~\ref{sec:twinhiggs}. 
\textbf{Right:} Constraints for $m_h=125$ GeV.  The three bands show the $1\sigma$ allowed regions: $R_{\gamma \gamma}$ (purple), $R_{b \bar b}$ (orange), $R_{\gamma \gamma jj}$ (beige).  The three curves show the theoretical predictions as a function of  $\xi$: $R_{\gamma \gamma}$ (solid-purple), $R_{b \bar b}$ (dashed-orange) and $R_{\gamma \gamma jj}$ (dotted-beige).  Only 3 channels are shown, but all channels are included. The region to the left of the green line at $\xi \simeq 0.6$ shows the 95\% CL experimental (combined) allowed region. }
 \label{TwinHiggsSigma}
}
\end{figure}

Another phenomenologically distinct example with one top partner arises within  the Twin Higgs scenario \cite{Chacko:2005pe}, where the global symmetry giving rise to a pseudo-Goldstone Higgs arises accidentally as a consequence of a discrete symmetry.
 In particular, in the left-right symmetric Twin Higgs model \cite{Chacko:2005un}
the top sector interactions with the Higgs take the form
\beq
-\cl_{top} =  y \sin(|H|/f) t T^c + y \cos(|H|/f) T t^c  + M_2 T T_c\,.
\eeq
Using the same methods as before  one finds,
\beq
c_V = c_b=   \sqrt{1 - {v^2 \over 2 f^2} }\,,
\qquad \qquad  {c_g \over c_{g,\rm SM}}  =  {c_\gamma \over c_{\gamma,\rm SM}}    = {1  - {v^2 \over f^2} \over  \sqrt{1 - {v^2 \over 2 f^2} } }\,.
\eeq
In this example, the  Higgs partial width into gluons is modified by a different factor than  that into W and Z bosons.
%Exactly the same formula is true, for example, in the $SO(5)/SO(4)$ minimal composite Higgs with fermions embedded in the fundamental representation of $SO(5)$ \cite{Agashe:2004rs}  (but in that case there is an  additional heavy quark in the top sector).    
The constraints on the non-universal suppression models are presented in Fig.~\ref{TwinHiggsSigma}. Assuming a 125 Gev Higgs, the allowed region at 95\% CL is $\xi< 0.6$.

%%%%%%%%%%%%%%%%%%%%%%%
\section{Multi-Higgs models}
\label{sec:multi-higgs}
%%%%%%%%%%%%%%%%%%%%%%%

%----------------------------------------------
\subsection{Doublet + Singlet}
\label{sec:ds}
\begin{figure}[tb]
\bc
\includegraphics[width=1\textwidth]{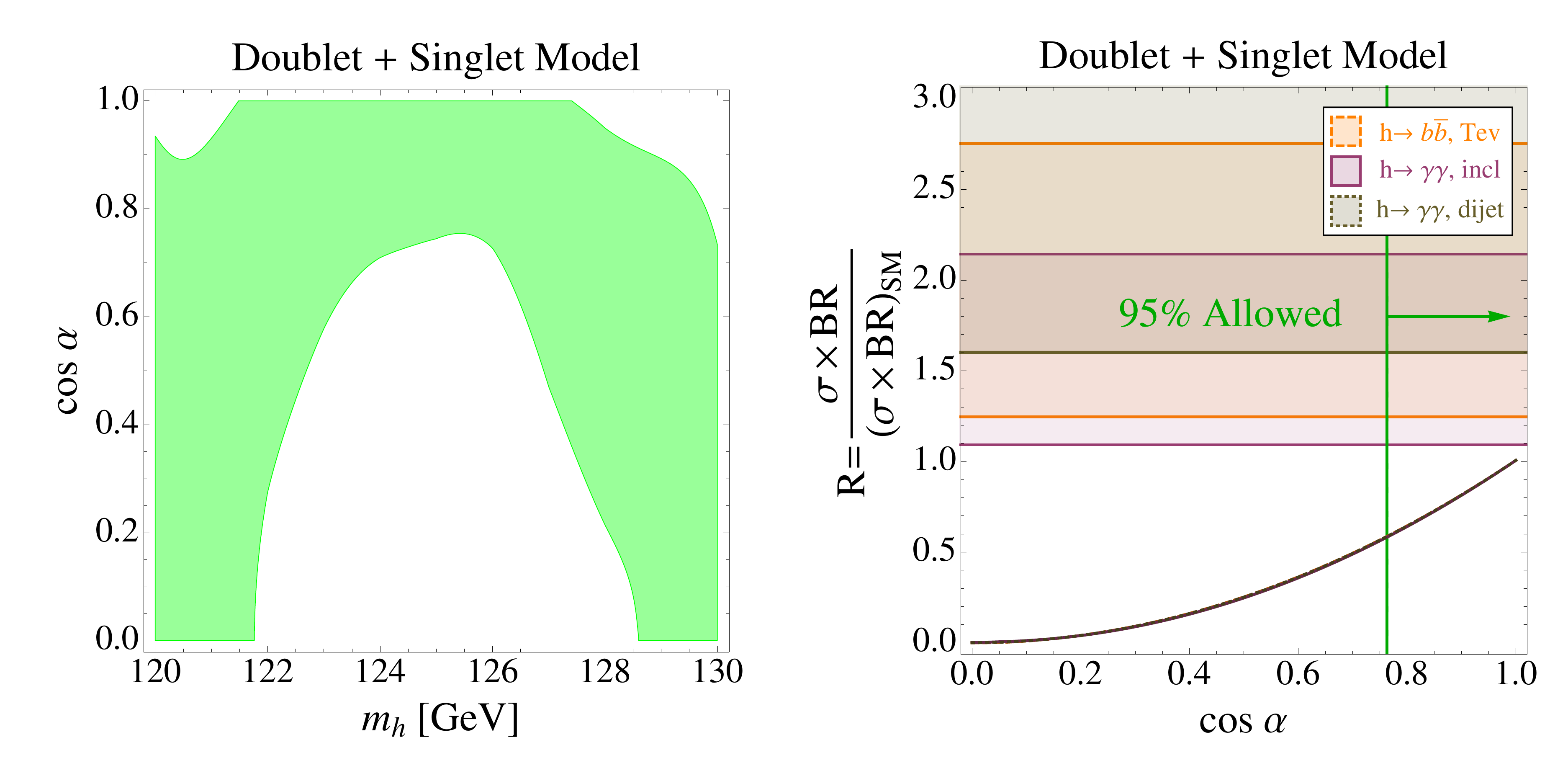}
\ec
\vspace{-1cm}
\caption{\small{\textbf{Left:} Favored region, 95\% CL, in the $\cos\alpha -m_h$ plane, derived from the combination of all search channels, for the doublet-singlet model described in Sec.~\ref{sec:ds}. 
\textbf{Right:} Constraints for $m_h=125$ GeV.  The three bands show the $1\sigma$ allowed regions: $R_{\gamma \gamma}$ (purple), $R_{b \bar b}$ (orange), $R_{\gamma \gamma jj}$ (beige). The three curves show the theoretical predictions as a function of  $\cos \alpha$: $R_{\gamma \gamma}$ (solid-purple), $R_{b \bar b}$ (dashed-orange) and $R_{\gamma \gamma jj}$ (dotted-beige).  Only 3 channels are shown, but all channels are included. The region to the right of the green line at $\cos \alpha \simeq 0.75$ shows the 95\% CL experimental (combined) allowed region. }
 \label{DoubletSingletR}}
\end{figure}

The simplest set-up with multiple Higgs bosons is the one with an electroweak-singlet scalar field mixing with the Higgs.
As a result, the mass eigenstates  are linear combinations of the Higgs scalar originating from the doublet (which couples to the SM matter) and singlet (which does not couple to matter).
Denoting the mixing angle as $\alpha$, all the couplings of the Higgs boson are suppressed by $\cos \alpha$,
\beq
c_V = c_b =  {c_g \over c_{g,\rm SM}}  =  {c_\gamma \over c_{\gamma,\rm SM}}    = \cos \alpha. 
\eeq
As a consequence, the Higgs production and decay rates in all the channels are universally suppressed by $\cos^2 \alpha$.
This is analogous to what happens in a fermionic model in Section \ref{s.slh}.
The new element is the appearance of the second Higgs eigenstate,  denoted by $H^0$, whose couplings are suppressed by $\sin \alpha$ compared to those of the SM Higgs boson, and whose mass is in general a free parameter.
In Fig.~\ref{DoubletSingletR} we present the LHC and Tevatron Higgs constraints on this model.  We find rather strong constraints on the mixing of the doublet with the singlet, $\cos\alpha \gtrsim 0.70$.    In deriving these constraints we assumed that $m_{H^0} > m_h/2$. 

%----------------------------------------------
\subsection{Two Higgs Doublets}

\begin{figure}[tb]
\bc
\includegraphics[width=0.45\textwidth]{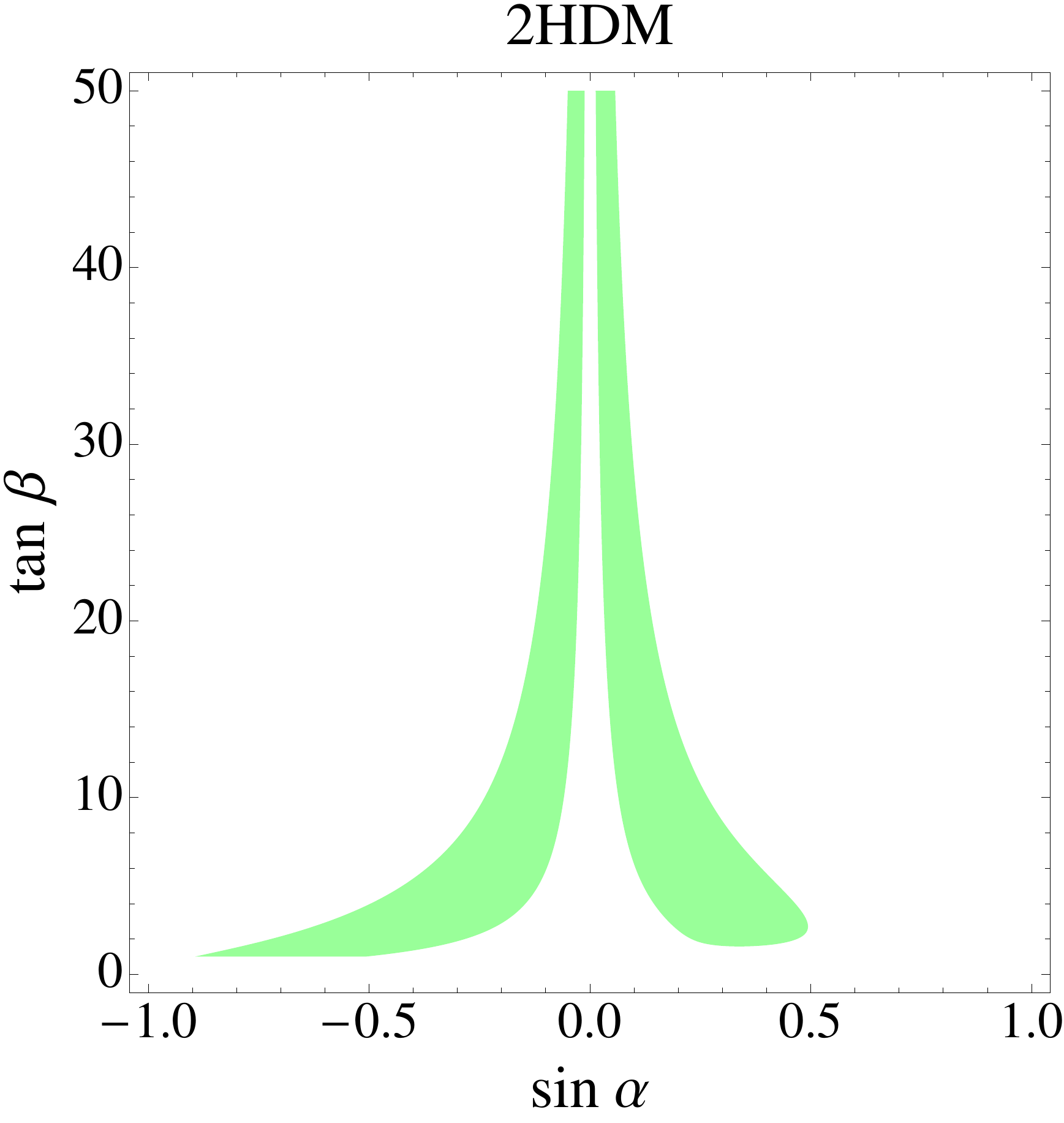}
\ec
\vspace{-1cm}
\caption{\small Favored region, 95\% CL,  of the 2HDM in the $\tan \beta-\sin \alpha$ plane, derived from the combination of all search channels. We take $m_{H^\pm}=200$ GeV, but a lighter charged Higgs would only slightly change the favored region. The favored region for $\sin \alpha < 0$  concentrates around the decoupling limit, $\alpha = \beta - \pi/2$, where all couplings are SM-like, whereas the region for $\sin \alpha >0$ lies around the region $\alpha = -\beta + \pi/2$ where the top Yukawa coupling is SM-like.  
 \label{2HDM}}
\end{figure}

We end with the study of 2 Higgs doublets $H_u$, $H_d$, the former coupling to up-type quarks, and the latter to down-type quarks and leptons.
The physical fields are embedded into the doublets as,
\begin{eqnarray}
H_u &=& \bvec \cos \beta H^+ \\ 
{1 \over \sqrt 2} ( v \sin \beta    +  h \cos \alpha + H^0 \sin \alpha  + i A^0 \cos \beta  )
\evec\,,
\\
H_d  &=& \bvec
{1 \over \sqrt 2} ( v \cos \beta   -   h \sin \alpha
+ H^0 \cos \alpha  + i A^0 \sin  \beta  )
\\
\sin\beta H^-
\evec\,.
\end{eqnarray}
The couplings of the lightest Higgs boson $h$ are described by two angle $\alpha$, $\beta$ who are in general free parameters\footnote{If the Higgs potential is that of the MSSM, the angle $\alpha$ is not independent of $m_{A^0}$ and $m_{H^\pm}$.  Furthermore, in that case $ -\pi/2 <  \alpha < 0$ for $m_{A^0}> m_Z$.}.   We find,
\beq
c_V  = \sin(\beta - \alpha)\,,
\quad
c_b = -{\sin \alpha \over \cos \beta}\,,
\quad
{c_g \over c_{g,\rm SM}}  =  {c_\gamma \over c_{\gamma,\rm SM}}    = {\cos \alpha \over \sin \beta}\,.
\eeq
By convention $0 < \beta < \pi/2$. In general, there is an additional contribution to $c_\gamma$ from the charged Higgs, but it is always small compared to the contribution from the W-boson. 

The 2 Higgs Doublet Model (2HDM) can change all couplings to the Higgs and thus is highly constrained by the LHC Higgs searches \cite{Ferreira:2011aa,Akeroyd:1998ui}. In Fig.~\ref{2HDM} we show the constraints in the $\tan \beta-\sin \alpha$ plane for $m_{H^\pm}=200$ GeV. Lighter masses would only slightly change the favored region. The favored region for $\sin \alpha < 0$ concentrates around the decoupling limit $\alpha = \beta - \pi/2$, where all couplings are SM-like. The region for $\sin \alpha >0$ lies around the region $\alpha = -\beta + \pi/2$ where the top Yukawa coupling is SM-like. 

%%%%%%%%%%%%%%%%%%%%%%%%%%%%%%%%%%%%%%%%%%%%
\section{Conclusions}
\label{sec:conclusions}

The indications for the existence of a Higgs boson provided recently by LHC and Tevatron are preliminary and may go away with more data.
With this caveat in mind, it is interesting to ask  whether the available experimental information is compatible with the SM Higgs boson, and whether it favors or disfavors any particular constructions beyond the SM.  In this paper we analyzed recent LHC and Tevatron searches sensitive to a light (115-130 GeV) Higgs boson, combining results in the channels: $h \to \gamma \gamma$ (both gluon fusion and vector-boson fusion), $h\to ZZ^* \to 4l$, $h\to WW^* \to 2l2\nu$, and $h \to b \bar b$, as well as combining the LHC and Tevatron data.
We presented interpretations  of that combination in the context of several effective models, with the special emphasis on models addressing the naturalness problem of electroweak symmetry breaking.

We have argued that, unsurprisingly,  the combination of the LHC and Tevatron data favors the Higgs boson in the mass range $124-126$ GeV, with the best fit cross section close to the one predicted by the SM.
Less trivially, we recast the LHC and Tevatron Higgs results as constraints on the parameters of the effective lagrangian at the scale $\sim m_h$ describing the leading interactions of the Higgs boson with the SM fields.
Furthermore, we found that the data  already put interesting constraints on simple natural new physics models, especially on those predicting suppression of  $\sigma(p p \to h ) {\rm Br} (h \to \gamma \gamma)$ and $\sigma(p p \to h j j) {\rm Br} (h \to \gamma \gamma)$.
For example, in a model with one fermionic top partner stabilizing the Higgs potential, the top partner masses below $\sim 220$ GeV are disfavored at 95\% CL.
For one scalar partner the corresponding bound is $\sim 240$ GeV, due to the fact that a single scalar stabilizing  the Higgs potential  always provides a positive contribution to $\Gamma(h \to \gamma \gamma)$.
These bounds can be further relaxed for more  complicated models.  In particular in a model with 2 scalar partners the total contribution to  $\Gamma(h \to \gamma \gamma)$ can be negligible even for very light scalars, at the expense of fine tuning.

We anticipate these bounds to significantly improve with additional data to be collected in 2012.   Alternatively, studying the effective theory of the Higgs bosons may prove to be the shortest way to a discovery of new physics beyond the SM.

\vspace{1cm}

{\bf Note added:} 
Right after our paper appeared, Refs.~\cite{Azatov:2012bz} and~\cite{Espinosa:2012ir} also appeared. 
The references also interpret the LHC Higgs results as constraints on the effective theory of Higgs interactions and overlap  in part with our work.
In order to assess compatibility with our results, in v2  we added the top right plot of Fig.~\ref{fig:generalFit}, which can be directly compared to the contours in the $a-c$ plane presented in~\cite{Azatov:2012bz,Espinosa:2012ir}.
In spite of using  different statistical methods, we find very similar preferred regions in the $a-c$  plane.   Nevertheless, our constraints on $\xi=v^2/f^2$ in Section~\ref{s.slh} are somewhat stronger than  in~\cite{Azatov:2012bz}.  
% as we have considered the reported results of the $ZZ^*$ and $\gamma\gamma$ channels separately,  where a hint for the Higgs seems the strongest.
We further note that our definition of $\xi$ differs by a factor of 2 compared to the definition in~\cite{Azatov:2012bz,Espinosa:2012ir}.

%%%%%%%%%%%%%%%
\section*{Acknowledgements}
%%%%%%%%%%%%%%
We especially thank Patrick Meade for collaboration in the early stages of this project and for many useful discussions.   We also thank David Curtin, Aielet Efrati, Yonit Hochberg, Yossi Nir and  Gilad Perez for useful discussions.   The work of DC, EK and TV is supported in part by a grant from the Israel Science Foundation.   The work of TV is further supported in part by the US-Israel Binational Science Foundation and the EU-FP7 Marie Curie, CIG fellowship.

%%%%%%%%%%%%%%%%%%%%%%%%%%%%%%%%%%%%%%%%%%%%%%%%%%%%%%%%%%%%%%%%%%

\end{document}